# Label-invariant models for the analysis of meta-epidemiological data

KM Rhodes[1], D Mawdsley[2], RM Turner[1,3], HE Jones[4], J Savović[4,5], JPT Higgins[4]

[1]MRC Biostatistics Unit, School of Clinical Medicine, University of Cambridge, Cambridge, UK
[2]University of Manchester, Manchester, UK
[3]MRC Clinical Trials Unit, University College London, UK
[4]School of Social and Community Medicine, University of Bristol, UK
[5]NIHR CLAHRC West, University Hospitals Bristol NHS Foundation Trust, Bristol, UK

*Abstract*

Rich meta-epidemiological data sets have been collected to explore associations between intervention effect estimates and study-level characteristics. Welton *et al.* proposed models for the analysis of meta-epidemiological data, but these models are restrictive because they force heterogeneity among studies with a particular characteristic to be at least as large as that among studies without the characteristic. In this paper we present alternative models that are invariant to the labels defining the two categories of studies. To exemplify the methods, we use a collection of meta-analyses in which the Cochrane Risk of Bias tool has been implemented. We first investigate the influence of small trial sample sizes (less than 100 participants), before investigating the influence of multiple methodological flaws (inadequate or unclear sequence generation, allocation concealment and blinding). We fit both the Welton *et al.* model and our proposed label-invariant model and compare the results. Estimates of mean bias associated with the trial characteristics and of between-trial variances are not very sensitive to the choice of model. Results from fitting a univariable model show that heterogeneity variance is, on average, 88% greater among trials with less than 100 participants. Based on a multivariable model, heterogeneity variance is, on average, 25% greater among trials with inadequate/unclear sequence generation, 51% greater among trials with inadequate/unclear blinding, and 23% lower among trials with inadequate/unclear allocation concealment, though the 95% intervals for these ratios are very wide. Our proposed label-invariant models for meta-epidemiological data analysis facilitate investigations of between-study heterogeneity attributable to certain study characteristics.

**Keywords:** meta-epidemiology; randomised trials; heterogeneity; Bayesian methods; Cochrane

## 1. Introduction

Meta-analysis is used to combine the results of multiple studies in order to synthesise evidence in a specific research area. Variation in effect sizes among studies, known as heterogeneity, is widespread and reflects differences in design and conduct of the studies, as well as differences in the characteristics of participants, interventions and outcomes studied. Possible explanations of heterogeneity among studies in a meta-analysis should be determined where possible. Meta-analysts may explore heterogeneity by separating studies into subgroups, using meta-regression or restricting analyses to studies with particular characteristics. However, results from such analyses are imprecise if the number of studies is small. On the other hand, combining all available studies, while ignoring differences in their design and conduct, may produce results that are difficult to interpret or, at worst, meaningless.

Meta-epidemiology is an emerging field of research that seeks to understand causes of heterogeneity across studies by re-analysing large numbers of meta-analyses. A notable example of a meta-epidemiological study is the *BRANDO* study [1], which combined data from several existing meta-epidemiological studies into a single database. Each of the 1973 included trials was categorised according to whether specific design characteristics were judged to be adequate, inadequate or unclear. Comparisons of trials in different categories were made within each of the meta-analyses, and these comparisons were combined across the 234 included meta-analyses. The results showed that, on average, effects were exaggerated in favour of the experimental treatment in trials judged not to have adequate sequence generation, allocation concealment and blinding.

The analysis of *BRANDO* followed methods proposed by Welton *et al.*[2], which model the biases that may arise due to particular design characteristics within a Bayesian framework. Specifically, a bias parameter is introduced for effect sizes in studies with a particular design characteristic such as inadequate allocation concealment, and a hierarchical model structure is assumed for these, with variability in bias assumed across the studies with the characteristic within each meta-analysis and across meta-analyses. While intuitively appealing, a limitation of these models is that they impose an additive relationship between the amount of heterogeneity in the groups of studies, such that the studies with the characteristic are constrained to be at least as heterogeneous as the studies without the characteristic [3]. The inherent assumption that studies without the characteristic will be no more heterogeneous than studies with the characteristic will not necessarily be true. Furthermore, when meta-epidemiological methods are used to examine study characteristics that are not clearly associated with methodological quality (such as single centre versus multicentre studies), the model is problematic because it forces a somewhat arbitrary assumption that the heterogeneity in one category is at least as large as that in the other category. In short, the model is not invariant to the labels defining the two categories of studies.

In this paper, we present general models for the analysis of meta-epidemiological studies that are label-invariant. We allow variation among the studies with the characteristic of interest to be higher or lower than the variation among the studies without the characteristic. We achieve this by modelling a multiplicative rather than an additive relationship between the variance components in the two categories of studies. After introducing this model for a univariable model for combining studies with and without a single reported characteristic, we generalise our approach to a multivariable model for combining studies that differ according to multiple characteristics. We then apply models with additive and label-invariant variance structures to a new meta-epidemiological dataset (the *ROBES* study), to investigate how estimates of average intervention effect and heterogeneity may depend on the choice of model. We assess the robustness of results from the proposed label-invariant model to the choice of prior distribution for the multiplicative parameter in section 6.

## 2. Univariable models to examine the influence of a single characteristic

We first describe models for meta-epidemiological studies involving a single characteristic, to investigate differences due to one particular attribute of the studies, such as an aspect of design or quality. In a given meta-analysis $m$, suppose that studies are categorised according to the presence or absence of the characteristic of interest. We assume that each study $i$ without the characteristic (denoted '−') provides an estimate of the underlying intervention effect $\theta_{im}^-$. We assume a normal random-effects distribution for the $\theta_{im}^-$ with mean $d_m$ and variance $\tau_m^2$, specific to meta-analysis $m$:

$$\theta_{im}^- \sim N(d_m, \tau_m^2).$$

A study with the characteristic (denoted '+') is assumed to estimate an underlying intervention effect $\theta_{im}^+$. In the additive model proposed by Welton *et al.*, we define this to be a potentially biased version of what would have been estimated in the absence of the characteristic:

$$\theta_{im}^+ = \theta_{im}^- + \beta_{im}.$$

A hierarchical structure is placed on the bias terms, $\beta_{im}$:

$$\beta_{im} \sim N(b_m, \kappa_m^2),$$

where $b_m$ and $\kappa_m^2$ represent the mean and variance in bias, within meta-analysis $m$, associated with presence of the characteristic. We use Model 3 of Welton *et al.*, assuming throughout that the variance in the biases is the same within each meta-analysis, such that $\kappa_m^2 = \kappa^2$ for all $m$. The assumption is not necessary, but there is seldom sufficient evidence to estimate multiple variance parameters. Thus the $\theta_{im}^+$ are assumed to be normally distributed as

$$\theta_{im}^+ \sim N(d_m + b_m, \tau_m^2 + \kappa^2).$$

Note that the consequence of the hierarchical model structure for the within-study biases is that the variance of the $\theta_{im}^+$ is constrained to be at least as high as the variance of the $\theta_{im}^-$. We will refer to this as an additive variance structure. This could be particularly problematic because the estimated variance components in Bayesian hierarchical models have been shown to be biased upwards [4]. For this reason, it may appear that the variance component $\kappa$ is strictly positive, even though it is not.

We propose an alternative, label-invariant, model, in which we allow the variance of the $\theta_{im}^+$ to be higher or lower than the variance of the $\theta_{im}^-$, by using a positive scale parameter, $\lambda$:

$$\theta_{im}^+ \sim N(d_m + b_m,\ \lambda \tau_m^2).$$

In the new model, $\lambda$ represents the ratio of the between-study heterogeneity among '+' studies with the characteristic of interest, compared with '−' studies without the characteristic. Thus, when $\lambda$ exceeds 1 the '+' studies show greater between-study variability than the '−' studies.

For both the Welton model and our proposed label-invariant model, the mean bias $b_m$ associated with the characteristic of interest in meta-analysis $m$ is assumed to be exchangeable across meta-analyses, with overall mean $b_0$ and between-meta-analysis variance in mean bias $\varphi^2$:

$$b_m \sim N(b_0,\ \varphi^2).$$

Note that because our model is invariant to whether the characteristic is present or absent, the $b_m$ and $b_0$ terms can be thought of simply as differences rather than as biases. Parameters $b_0$, $\varphi$ and $\kappa$ or $\lambda$ are estimated through fitting these models to a meta-epidemiological dataset.

### 3. Multivariable models to examine the influence of multiple characteristics

Suppose now that studies in each pair-wise meta-analysis $m$ have been categorised according to the presence of $p$ reported characteristics, again representing differences in design or quality. We set the indicator $X_{ijm}$ to be 1 for studies with the $j$-th reported characteristic ($j=1,2,\ldots,p$), and 0 for studies without that characteristic.

Multivariable models are based on extensions of the hierarchical models described in section 2. In the additive multivariable model, the studies with characteristic $j$ (studies with $X_{ijm}=1$) are assumed to estimate the same underlying intervention effect as the studies without this characteristic (studies with $X_{ijm}=0$) plus some study-specific, characteristic-specific bias $\beta_{ijm}$. In multivariable analyses presented for the *BRANDO* study [1], a generalised version of Welton *et al*'s model was fitted, assuming each study $i$ to estimate an underlying intervention effect $\theta_{im}$:

$$\theta_{im} \mid X_{1im},...,X_{pim}, \beta_{1im},...\beta_{pim} \sim N\left(d_m + \sum_{j=1}^{p} X_{ijm}\beta_{ijm}, \tau_m^2\right).$$

That is,

$$\theta_{im} \mid X_{1im},...,X_{pim} \sim N\left(d_m + \sum_{j=1}^{p} X_{ijm}b_{jm}, \tau_m^2 + \sum_{j=1}^{p} \kappa_j^2\right)$$

In the label-invariant model, the studies are assumed to have underlying intervention effect:

$$\theta_{im} \mid X_{1im},...,X_{pim} \sim N(d_m + \sum_{j=1}^{p} X_{ijm}b_{jm},\ \tau_m^2 \prod_{j=1}^{p} \{(1-X_{ijm}) + X_{ijm}\lambda_j\}).$$

For example, the intervention effect $\theta_{im}$ in a study $i$ with characteristics 1 and 2 but not 3 would have a normal distribution with mean $d_m + b_{1m} + b_{2m}$ and variance $\tau_m^2 \lambda_1 \lambda_2$.

Under this model, we estimate the parameters representing the differences $b_m$ in between-study means and the ratios $\lambda_j$ of between-study variances, for each of the $p$ characteristics.

For both models, the implied average bias (on the log odds ratio scale) in studies with any combination of study characteristics is estimated by the sum of the relevant fitted $b_0$ terms $\sum_j b_{0j}$. In practice we might expect study characteristics to be correlated. It would be possible to extend the multivariable models to include interactions between different study characteristics [1], but we do not explore interactions in this paper.

## 4. Additional modelling of heterogeneity variance

It is of interest to compare variance estimates from fitting the existing additive model and the proposed label-invariant model. This is difficult because estimates of $\kappa$ from the additive model and $\lambda$ from the label-invariant model have different interpretations and are not directly comparable. For this reason, we focus on estimates of heterogeneity variances $\tau_m^2$ (among studies without the characteristic of interest) which have the same interpretation under the two different models. These estimates are influenced by $\kappa$ in the additive model and $\lambda$ in the label-invariant model, and hence comparison of the distributions obtained under the two different models gives some indication of agreement between estimates of variance components.

Between-study heterogeneity $\tau_m^2$ varies across meta-analyses *m* and is estimated in the additive and label-invariant models above. However, in order to compare estimates of total (within meta-analysis) heterogeneity, we need to obtain some "typical" value of $\tau$. For this reason, we extend the models slightly to include a hierarchical structure for log($\tau_m^2$). We follow the approach of Turner *et al.* [5] and model the underlying values of between-study variance $\tau_m^2$ in intervention effect among studies without the characteristic of interest, assuming these to follow a log-normal($\mu, \sigma^2$) distribution. A predictive distribution for the heterogeneity variance $\tau_{new}^2$ (among studies without the characteristic of interest) is obtained under the full Bayesian model:

$$\tau_{new}^2 \sim \text{log-normal}\ (\mu, \sigma^2).$$

To summarise this distribution, we report a log-normal distribution fitted to the predictive distribution, using the posterior mean and standard deviation for log($\tau_{new}^2$).

## 5. Applications

We make use of a newly constructed database from the *ROBES* study [6]. This database comprises 244 meta-analyses with completed Risk of Bias tables, extracted from the April 2011 issue of the *Cochrane Database of Systematic Reviews* (*CDSR*). For each trial in every meta-analysis, information is available on all items addressed by the Cochrane Risk of Bias tool [7]. Meta-analyses where numerical data were unavailable or where pooling was considered inappropriate were excluded. The *ROBES* database also excluded meta-analyses comprising fewer than five trials. One binary outcome meta-analysis from each eligible Cochrane review was included in the database, corresponding to a primary outcome where possible.

Empirical evidence suggests that intervention effect estimates are exaggerated in smaller studies [8] and studies with flaws in their design and conduct [1;6]. As an example application of the univariable models, we investigate the influence of small study sample sizes (less than 100 participants), before using multivariable models to investigate the influence of multiple methodological flaws (inadequate or unclear sequence generation, allocation concealment and blinding).

In all analyses, intervention effects were modelled as log odds ratios and outcomes were coded so that a log odds ratio less than 0 corresponded to a beneficial intervention effect. We assumed the observed number of events in each arm of each trial to have a binomial distribution. A vague normal(0,1000) prior was assigned to all location parameters. Estimated mean differences $b_0$ in intervention effect between trials with and without the characteristic of interest were exponentiated and are therefore reported as relative odds ratios (ROR). Meta-analyses may be informative for some, but not all, reported characteristics. We refer to meta-analyses as informative for a characteristic if they contained at least one trial with the characteristic and one without the characteristic and could inform estimation of average differences in intervention effect $b_m$ associated with the characteristic.

In the additive models, variance parameters $\kappa^2$ and $\varphi^2$ were given modified inverse-gamma(0.001,0.001) prior distributions with a probability atom at zero variance, following Savovic *et al.* [1]. That is, we let the variance parameters $\kappa^2$ and $\varphi^2$ be equal to zero with some probability $p_0$ and equal to the variance from an inverse-gamma prior with probability $(1-p_0)$. The mixing probability $p_0$ was given an uninformative Beta(1,1) prior. For the additional modelling of heterogeneity variances, we placed a vague prior on the mean, $\mu$, and assumed a uniform(0,2) prior for the standard deviation parameter σ, representing variation in heterogeneity across meta-analyses, as suggested by Spiegelhalter *et al.* [9]. Posterior summaries from all models were obtained by using Markov chain Monte Carlo (MCMC) methods within *WinBUGS* Version 1.4.3 [10]. In order to produce very low MC error rates, we based results on 500,000 iterations, following a burn-in period of 25,000 iterations which was sufficient to achieve convergence. Convergence was assessed according to the Brooks-Gelman-Rubin diagnostic tool [11], with three chains starting from widely dispersed initial values.

In the label-invariant models, we placed a log-normal(0,1) prior distribution on the multiplicative parameters $\lambda$, which has median 1 on the untransformed scale. We assess sensitivity to this choice of prior distribution in section 6. Since it was not possible to compute both $\tau_m$ and $\kappa$ or $\lambda$ in a meta-analysis with fewer than two trials with and without a characteristic of interest, such meta-analyses were not allowed to contribute to the estimation of $\kappa$ in the additive models or to the estimation of $\lambda$ in the label-invariant models through use of the "cut" function in *WinBUGS*.

In multivariable analyses we used *WinBUGS* to calculate a 95% credible interval for the implied average bias in trials with any combination of study characteristics, $\sum_j b_{0j}$, that accounted for correlations between the coefficients $b_{0j}$.

Model fit comparison was based on the deviance information criterion (*DIC*) [12;13]. The posterior mean of the total residual deviance $D_{res}$ was used to assess the goodness-of-fit of the hierarchical models. The *DIC* provides a measure of model fit that penalizes $D_{res}$ by the effective number of parameters $p_D$. Due to the non-linearity between the likelihood and the model parameters, we calculated $p_D$ at the posterior mean of the fitted values rather than at the posterior mean of the parameters [14].

The WinBUGS code for the label-invariant models presented in the paper is available in Supporting Information.

**5.1 Example 1: Univariable analyses examining the influence of sample size less than 100**

Sample size can vary substantially among studies, even within a single meta-analysis addressing the same research question [15]. Dechartres *et al.* investigated the influence of trial sample size on treatment effect estimates in a large collection of meta-analyses of various medical conditions and interventions [8]. Effect estimates differed within meta-analyses according to trial sample size; on average, stronger estimates were observed in small to moderately sized studies than in the largest studies. Here we apply univariable models to examine the influence of trial sample sizes of less than 100 participants on intervention effect and between-trial heterogeneity. This fixed threshold for sample size has been chosen because it is approximately the median sample size of trials included in the *ROBES* database; our intention is to exemplify the methods rather than to provide empirical evidence.

We analysed data from 2091 trials included in 179 binary outcome meta-analyses that were informative to detect differences in intervention effect between trials with sample size less than 100 participants and those with larger sample sizes. The number of trials per meta-analysis ranges from 5 to 75 with median 9 and IQR 6 to 13. The number of trials per meta-analysis with less than 100 participants ranges from 1 to 29 with median 4 and IQR 2 to 7.

Posterior summaries derived from the additive model are reported in Table 1. Intervention effects were exaggerated by an average of 15% in trials with sample sizes of less than 100 (ROR 0.85, 95% credible interval (CI): 0.76 to 0.93). The posterior median of the between-trial within-meta-analysis standard deviation $\kappa$ is 0.22 (95% CI: 0.02 to 0.40), and the between-meta-analysis standard deviation in mean bias $\varphi$ has posterior median 0.22 (95% CI: 0.03 to 0.36). We derived a predictive log-normal(-2.94, $1.69^2$) distribution for between-trial variance $\tau^2_{new}$ expected among trials with a sample size of at least 100 participants. This predictive distribution has median 0.05 and 95% range 0.002 to 1.42.

Table 1 also provides the posterior summaries obtained from the label-invariant model. Estimated mean bias is fairly robust to the choice of variance structure, as is the estimated between-meta-analysis standard deviation in mean bias $\varphi$. In the label-invariant model, $\lambda$ has posterior median 1.88 (95% CI: 1.08 to 3.10), indicating that variation among trials with sample sizes less than 100 is on average 88% greater than that among trials with at least 100 participants. The predictive distribution obtained for the between-trial heterogeneity variance $\tau^2_{new}$ expected among trials with at least 100 participants is fairly similar to that obtained under the additive model, but gives less support to higher levels of heterogeneity.

Values of $D_{res}$ and $p_D$ for the additive model are close to those for the label-invariant model (Table 1). The label-invariant model has a lower $DIC$, but the difference in $DIC$ between the additive and label-invariant models is less than 5 and hence not considered to be meaningful [12]. We therefore have little reason to choose one model over the other.

To confirm the label-invariant property of our proposed model, we inverted the sample size labels of the trials and applied the univariable models to examine the influence of trial sample sizes greater than 100 participants on intervention effect and between-trial heterogeneity. Parameter estimates are shown in Table 2. As expected, estimated mean bias $b_0$ is the negative of the estimate obtained before re-labelling the trials. Under the additive model, between-trial heterogeneity (quantified by $\kappa$) is constrained to be higher for trials with greater than 100 participants; the posterior median of $\kappa$ is 0.03 (95% CI: 0.01 to 0.09), which is lower than the posterior median 0.22 in Table 1. Under the label-invariant model, the posterior median of $\lambda$ is 0.53 (95% CI: 0.32 to 0.85), which is approximately the reciprocal of the posterior median 1.88 obtained before relabelling the trials. After re-labelling the trials, we find that the label-invariant model gives a better fit (lower $D_{res}$ and $DIC$) and is preferred over the additive model.

Although the same label-invariant model is used in each analysis reported in Tables 1 and 2, we make different assumptions about the distributional form of heterogeneity variances by relabelling the trials. For example, in the first analysis comparing smaller vs. larger studies (Table 1), heterogeneity variance for larger trials with greater than 100 participants is $\tau^2$, which is assumed to have a log-normal distribution. After re-labelling the trials, heterogeneity variance among larger trials is $\lambda\tau^2+\varphi^2$, which is not log-normal. As a consequence, model fit may differ between the two analyses; here, we find that the label-invariant model gives a better fit (lower $D_{res}$ and $DIC$) after re-labelling the trials.

**5.2 Example 2: Multivariable analyses examining the influence of inadequate or unclear sequence generation, allocation concealment and blinding.**

When conducting a systematic review, it is important to consider the risk of bias in the results of included studies. Including biased studies with methodological flaws in a meta-analysis will cause the results of the meta-analysis to be biased. Differences in risks of bias can help explain variation in the results of the studies. We conducted multivariable analyses to examine the influences of inadequate or unclear sequence generation, allocation concealment and blinding, compared with adequate, on combined intervention effect and between-trial heterogeneity. Table 3 presents results from these analyses, based on 117 informative meta-analyses (1473 trials) in which all three design characteristics were assessed. The number of trials per meta-analysis ranges from 5 to 75 with median 10 and IQR 6 to 14. In 396 (27%) trials, all three design characteristics were judged as inadequate or unclear. All three design characteristics were assessed as adequate in 361 (24%) trials.

Posterior summaries from the multivariable additive model are shown in Table 3. After adjusting for allocation concealment and blinding, intervention effects were exaggerated by 5% on average in trials with inadequate or unclear sequence generation (ROR 0.95, 95% CI: 0.87 to 1.04). There is evidence that between-trial heterogeneity (quantified by $\kappa_1$) is increased for trials with inadequate or unclear sequence generation; the posterior median of the additional between-trial within-meta-analysis standard deviation $\kappa_1$ is 0.12 (95% CI: 0.02 to 0.26). For inadequate or unclear allocation concealment, the ROR has posterior median 0.96 with standard deviation 0.04, while $\kappa_2$ has posterior median 0.06 (95% CI: 0.01 to 0.20) after adjustment for sequence generation and blinding. Inadequate or unclear blinding was associated with an average 8% exaggeration of intervention effect estimates (ROR 0.92, 95% CI: 0.85 to 0.99) and with increased heterogeneity ($\kappa_3$ 0.08, 95% CI: 0.01 to 0.27) after adjustment for sequence generation and allocation concealment. The posterior estimates for mean bias imply an average bias of 0.84 (95% range: 0.76 to 0.92), on the relative odds ratio scale, for a trial judged as inadequate or unclear for all three bias domains.

Table 3 also gives the posterior summaries obtained under the proposed label-invariant model. Estimates for mean bias $b_{0j}$ are not very sensitive to the choice of variance structure. We find that estimates of between-meta-analysis standard deviation $\varphi_j$ in mean bias are comparable to those obtained using the additive model.

In the label-invariant model, the posterior median for $\lambda_1$ indicates that variation among trials with inadequate or unclear sequence generation is on average 25% greater than that among trials with adequate sequence generation. Heterogeneity among trials judged as inadequate or unclear for allocation concealment is on average 77% of that among trials assessed as low risk of bias due to allocation concealment. The central estimate for $\lambda_3$ suggests that variation among trials with inadequate or unclear blinding is on average 51% greater than that among trials with adequate blinding. However, we note that each $\lambda_j$ is imprecisely estimated in the multivariable label-invariant model; the 95% credible intervals for $\lambda_j$ are wide and contain the null value 1 representing no difference.

The predictive distributions obtained for the between-trial variance $\tau^2_{new}$ expected to remain after "removing" bias due to inadequate or unclear sequence generation; allocation concealment and blinding are similar under the two models.

In this example, we find that the additive model gives a better fit (lower $D_{res}$ and $DIC$) and is preferred over the label-invariant model.

## 6. Assessing the sensitivity of results to the choice of prior for the multiplicative parameter λ

Here we assess the sensitivity of posterior inferences from the label-invariant meta-epidemiological analyses to the choice of prior distribution for $\lambda_j$. We re-analysed data from the 117 meta-analyses which were informative for bias due to inadequate/unclear sequence generation, allocation concealment and blinding. We examined five realistic candidate vague priors for parameters $\lambda_j$ (Figure 1). Priors 1 to 4 are centred about the null value 1 representing no difference in heterogeneity among trials judged as adequate for all three bias domains and trials judged as inadequate or unclear for design characteristic $j$. Prior 5 (truncated log-normal) gives support to values of $\lambda_j$ strictly greater than 1, therefore mimicking the existing additive model for meta-epidemiological data analysis. This prior has median 1.96 and 95% range 1.03 to 9.40. Although each prior distribution is considered to be vague, each represents different beliefs about the likely ratio by which between-trial heterogeneity changes for trials with inadequate/unclear design characteristics. For example, a uniform prior distribution for log($\lambda_j$) assigns equal weight to all values in the chosen range, giving no preference to certain values. In contrast, a normal prior distribution for log($\lambda_j$) gives greater support to values of log($\lambda_j$) closer to the centre of the distribution, therefore giving less weight to the lower and higher values in its range.

For bias due to inadequate or unclear blinding, differences between posterior estimates from fitting models with the five different priors are illustrated in Figure 1. Comparable results were obtained for bias due to inadequate or unclear sequence generation and allocation concealment (results not shown). The posterior distributions for mean bias $b_0$ and between meta-analysis variance in mean bias $\varphi$ are fairly robust to the choice of prior for the multiplicative parameter $\lambda_j$ in the model. Posterior estimates for the multiplicative parameter $\lambda_j$ are somewhat consistent among the different priors, with overlapping credible intervals, but the more precise priors 1, 3 and 5 unsurprisingly lead to lower posterior standard deviations for $\lambda_3$.

## 7. Discussion

We have proposed univariable and multivariable label-invariant models for conducting meta-epidemiological analyses to investigate the influence of a single study characteristic or multiple study characteristics on intervention effect and heterogeneity in a meta-analysis. The label-invariant models are modified versions of a model proposed by Welton *et al.* [2]. When considering heterogeneity among effect sizes, our methods allow us to distinguish between variation due to known study characteristics and other sources of between-study variation, as recommended by Higgins *et al.* [16]. Our label-invariant models are more flexible than the models of Welton *et al.*, in allowing us to quantify the ratio by which between-study heterogeneity changes for studies with certain characteristics.

We applied the existing additive and proposed label-invariant univariable and multivariable models to the *ROBES* [6] database. As an example application of the univariable models, we investigated the influence of trial sample sizes of less than 100 participants. As an example application of the multivariable models, we investigated the influence of inadequate or unclear sequence generation, allocation concealment and blinding. The findings in these examples give little to choose between the additive and label-invariant models; differences between the two models in deviance information criterion were small and not very meaningful. Reassuringly for the meta-epidemiological studies that use the additive model of Welton *et al.* [2], estimates of mean and variance parameters in the model were not very sensitive to the way that the meta-epidemiological data were modelled. However, the additive model gave inconsistent results; results for the influence of trial sample size less than 100 participants showed increased heterogeneity among smaller studies, but after re-labelling the trials to investigate the influence of trial sample size at least 100 participants results showed increased heterogeneity among larger studies (95% credible intervals did not contain the null value 0). We would therefore propose using the label-invariant model in future meta-epidemiological studies, on the grounds that the additive model is less general, and that it would be reasonable to allow heterogeneity among studies with a certain characteristic to be higher or lower than that among studies without the characteristic.

Empirical studies have investigated the impact of small studies on meta-analysis results [8;17;18]. There is empirical evidence to suggest that estimates of intervention effect are exaggerated in smaller studies, but little attention has been paid to small study effects on heterogeneity. Through meta-epidemiological analyses of the *ROBES* dataset, we have obtained empirical evidence that trials with less than 100 participants are more heterogeneous. Based on results from fitting our proposed label-invariant model, variation among trials with sample size less than 100 is, on average, 88% greater than that among trials with at least 100 participants. Exploring estimated heterogeneity among small studies should form the subject of future work. In particular, we have focussed on a dataset comprising meta-analyses of binary outcome data, and our results cannot necessarily be generalised to meta-analyses of continuous or other outcomes. Results of meta-epidemiological studies using different types of outcome data would be of interest.

In future work we plan to conduct an empirical study of the extent of heterogeneity in a meta-analysis that is due to within-study biases. Our planned empirical study will contribute to our understanding of associations between study-level characteristics and the extent of heterogeneity in a meta-analysis. It might occur that a trial characteristic is associated with bias in intervention effect, but that the bias does not explain between-trial heterogeneity in the meta-analysis. In this situation, using the formulae of Welton et al. to correct for bias in a new meta-analysis may inappropriately lead to greater down-weighting of the trials with the characteristic, whereas our label-invariant models may give the trials with the characteristic relatively more weight.

In the example applications, we found that the credible intervals obtained for the multiplicative parameter $\lambda$ in our label-invariant model were typically very wide. This is of concern because the estimates of $\lambda$ are of substantial interest, quantifying the ratio by which between-study heterogeneity increases for studies with the characteristic of interest. We advise caution in interpreting estimates of $\lambda$, in particular those close to 1 and those for which the lower or upper bound of the credible interval is close to 1. We suggest the use of non-informative priors that exclude implausibly high values in analyses of meta-epidemiological data. In any meta-epidemiological analysis, we recommend assessing the sensitivity of results to the choice of prior distributions for unknown parameters, particularly when data comprise meta-analyses of only a small number of studies (e.g. fewer than 10).

In summary, we have proposed label-invariant models for meta-epidemiological analyses investigating the influence of a single study characteristic or multiple study characteristics on intervention effect and heterogeneity. Unlike existing methods, our approach does not constrain between-study heterogeneity to be higher for studies with specific characteristics than for studies without the characteristics, and thus facilitates empirical investigations of heterogeneity.

**Acknowledgements**

KR and RT were supported by UK Medical Research Council (MRC) grants MR/K014587/1 and U105260558. DM was supported by a National Institute of Health Research (NIHR) Research Methods Fellowship and an MRC industry collaboration agreement, which was part funded by Pfizer. HEJ was supported by a MRC career development award in biostatistics (MR/M014533/1). JS was supported by the MRC Fellowship (Savović; G0701659/1) and by the NIHR Collaboration for Leadership in Applied Health Research and Care West (CLAHRC West) at University Hospitals Bristol NHS Foundation Trust. The views expressed are those of the authors and not necessarily those of the MRC, the National Health Service, the NIHR or the Department of Health.

**Table 1.** Posterior summaries from univariable models with additive and label-invariant variance structures, examining the influence of sample size less than 100 participants.

| | Additive model | | | Label-invariant model | | |
|---|---|---|---|---|---|---|
| **Parameters in model** | **Median** | **SD** | **95% CI** | **Median** | **SD** | **95% CI** |
| $b_0$ | -0.16 | 0.05 | -0.25 to -0.07 | -0.17 | 0.04 | -0.25 to -0.09 |
| ROR | 0.85 | 0.04 | 0.76 to 0.93 | 0.84 | 0.04 | 0.78 to 0.92 |
| $\lambda$ | n/a | | | 1.88 | 0.52 | 1.08 to 3.10 |
| $\kappa$ | 0.22 | 0.11 | 0.02 to 0.40 | n/a | | |
| $\varphi$ | 0.22 | 0.09 | 0.03 to 0.36 | 0.19 | 0.09 | 0.03 to 0.36 |
| **Predictive distributions for heterogeneity among trials with sample size ≥ 100 in a new meta-analysis** | | | | | | |
| $\tau^2_{new}$ (heterogeneity among trials with $X_{im}$=0) | Log-normal(-2.94, 1.69$^2$) Median=0.05, 95% range 0.002 to 1.42 | | | Log-normal(-2.96, 1.59$^2$) Median=0.05, 95% range 0.002 to 1.13 | | |
| Model fit | $D_{res}$=4183, $p_D$=2838, DIC=7021 | | | $D_{res}$=4186, $p_D$=2832, DIC=7018 | | |

SD standard deviation; CI credible interval; $b_0$ average difference in intervention log odds ratio associated with smaller sample sizes, ROR average change in estimated intervention effects for smaller trials (ratio of odds ratios); $\kappa$ average increase in between-trial heterogeneity for smaller trials within meta-analyses; $\lambda$ average change in heterogeneity variance for smaller trials; $\varphi$ between-meta-analysis variance in the average difference in intervention effect associated with smaller sample sizes; $D_{res}$ posterior mean of the total residual deviance; $p_D$ effective number of parameters; *DIC* deviance information criterion.

**Table 2.** Posterior summaries from the univariable model with label-invariant variance structure, examining the influence of sample size greater than 100 participants.

| | Additive model | | | Label-invariant model | | |
|---|---|---|---|---|---|---|
| **Parameters in model** | **Median** | **SD** | **95% CI** | **Median** | **SD** | **95% CI** |
| $b_0$ | 0.16 | 0.04 | 0.08 to 0.23 | 0.17 | 0.04 | 0.09 to 0.25 |
| ROR | 1.17 | 0.04 | 1.08 to 1.25 | 1.18 | 0.05 | 1.09 to 1.28 |
| $\lambda$ | n/a | | | 0.53 | 0.13 | 0.32 to 0.85 |
| $\kappa$ | 0.03 | 0.02 | 0.01 to 0.09 | n/a | | |
| $\varphi$ | 0.24 | 0.07 | 0.03 to 0.34 | 0.20 | 0.09 | 0.02 to 0.35 |
| Model fit | $D_{res}$=4208, $p_D$=2813, DIC=7021 | | | $D_{res}$=4177, $p_D$=2572, DIC=6749 | | |

SD standard deviation; CI credible interval; $b_0$ average difference in intervention log odds ratio associated with larger sample sizes, ROR average change in estimated intervention effects for larger trials (ratio of odds ratios); $\kappa$ average increase in between-trial heterogeneity for larger trials within meta-analyses; $\lambda$ average change in heterogeneity variance for larger trials; $\varphi$ between-meta-analysis variance in the average difference in intervention effect associated with larger sample sizes; $D_{res}$ posterior mean of the total residual deviance; $p_D$ effective number of parameters; *DIC* deviance information criterion.

**Table 3**. Posterior summaries from the multivariable models with additive and label-invariant variance structures, examining the influence of inadequate or unclear sequence generation, allocation concealment and blinding.

|  | Additive model | | | Label-invariant model | | |
|---|---|---|---|---|---|---|
| **Parameters in model** | **Median** | **SD** | **95% CI** | **Median** | **SD** | **95% CI** |
| **Inadequate or unclear sequence generation** | | | | | | |
| $b_{01}$ | -0.05 | 0.05 | -0.14 to 0.04 | -0.04 | 0.04 | -0.13 to 0.05 |
| ROR | 0.95 | 0.04 | 0.87 to 1.04 | 0.96 | 0.04 | 0.88 to 1.05 |
| $\lambda_1$ | n/a | | | 1.25 | 0.55 | 0.56 to 2.71 |
| $\kappa_1$ | 0.12 | 0.07 | 0.02 to 0.26 | n/a | | |
| $\varphi_1$ | 0.14 | 0.07 | 0.02 to 0.29 | 0.15 | 0.07 | 0.02 to 0.29 |
| **Inadequate or unclear allocation concealment** | | | | | | |
| $b_{02}$ | -0.04 | 0.04 | -0.12 to 0.04 | -0.04 | 0.04 | -0.12 to 0.04 |
| ROR | 0.96 | 0.04 | 0.88 to 1.05 | 0.96 | 0.04 | 0.89 to 1.04 |
| $\lambda_2$ | n/a | | | 0.77 | 0.42 | 0.33 to 1.93 |
| $\kappa_2$ | 0.06 | 0.05 | 0.01 to 0.20 | n/a | | |
| $\varphi_2$ | 0.06 | 0.05 | 0.01 to 0.19 | 0.06 | 0.05 | 0.01 to 0.21 |
| **Inadequate or unclear blinding** | | | | | | |
| $b_{03}$ | -0.09 | 0.04 | -0.17 to -0.01 | -0.09 | 0.04 | -0.17 to -0.01 |
| ROR | 0.92 | 0.04 | 0.85 to 0.99 | 0.92 | 0.04 | 0.85 to 0.99 |
| $\lambda_3$ | n/a | | | 1.51 | 0.63 | 0.71 to 3.14 |
| $\kappa_3$ | 0.08 | 0.07 | 0.01 to 0.27 | n/a | | |
| $\varphi_3$ | 0.09 | 0.06 | 0.01 to 0.23 | 0.11 | 0.07 | 0.01 to 0.25 |
| **Implied average bias in studies judged as inadequate/unclear for all three design characteristics** | | | | | | |
| $b_0$ | -0.18 | 0.05 | -0.28 to -0.08 | -0.17 | 0.05 | -0.27 to -0.07 |
| ROR | 0.84 | 0.04 | 0.76 to 0.92 | 0.85 | 0.04 | 0.76 to 0.94 |
| **Predictive distributions for heterogeneity remaining after "removing" bias in a new meta-analysis** | | | | | | |
| $\tau^2_{new}$ (heterogeneity among trials with $X_{im}$=0) | Log-normal(-3.89,1.84$^2$) Median=0.02, 95% range <0.001 to 0.72 | | | Log-normal(-3.89,1.79$^2$) Median=0.02, 95% range <0.001 to 0.65 | | |
| Model fit | $D_{res}$=2962, $p_D$=1922, DIC=4884 | | | $D_{res}$=2978, $p_D$=1913, DIC=4891 | | |

SD standard deviation; CI credible interval; $b_0$ average difference in intervention log odds ratio associated with the characteristic; ROR average change in estimated intervention effects for trials with the characteristic (ratio of odds ratios); $\kappa$ average increase in between-trial heterogeneity for trials with the characteristic within meta-analyses; $\lambda$ average change in heterogeneity variance for trials with the characteristic of interest; $\varphi$ between-meta-analysis variance in the average difference in intervention effect associated with the characteristic; $D_{res}$ posterior mean of the total residual deviance; $p_D$ effective number of parameters; DIC deviance information criterion.

**Figure 1.** Posterior medians and 95% credible intervals for mean bias due to inadequate/unclear blinding $b_{03}$, between meta-analysis standard deviation in mean bias $\varphi_3$, and the multiplicative parameter $\lambda_3$.

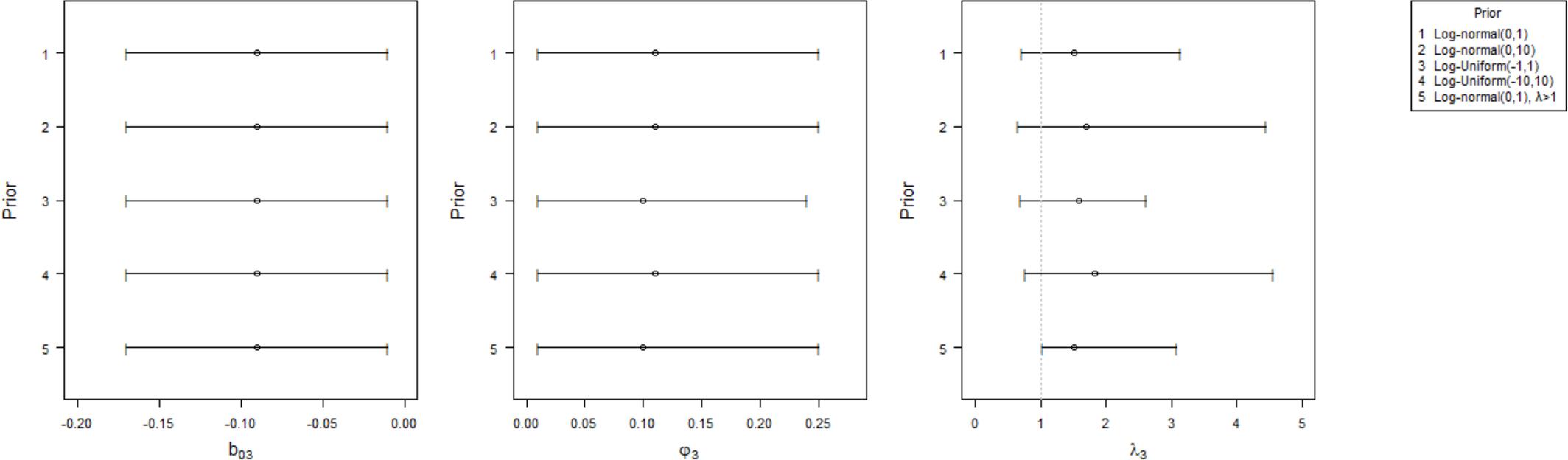